\documentclass{PoS}
\usepackage{amsmath}
\def\bra#1{\ensuremath{\langle{#1}\vert}}
\def\ket#1{\ensuremath{\vert{#1}\rangle}}

\title{$\eta$-$\eta'$ mixing in large-$N_c$ chiral perturbation theory: discussion, phenomenology, and prospects}

\ShortTitle{$\eta$-$\eta'$ mixing in L$N_c$ChPT: discussion, phenomenology, and prospects}

\author{Patricia Bickert\\
        PRISMA Cluster of Excellence, Institut f\"ur Kernphysik, Johannes Gutenberg-Universit\"at Mainz,
D-55099 Mainz, Germany\\
        E-mail: \email{bickert@kph.uni-mainz.de}}

\author{\speaker{Pere Masjuan}%

\\
        PRISMA Cluster of Excellence, Institut f\"ur Kernphysik, Johannes Gutenberg-Universit\"at Mainz,
D-55099 Mainz, Germany\\
       E-mail: \email{masjuan@kph.uni-mainz.de}}

\author{Stefan Scherer\\
        PRISMA Cluster of Excellence, Institut f\"ur Kernphysik, Johannes Gutenberg-Universit\"at Mainz,
D-55099 Mainz, Germany\\
        E-mail: \email{scherer@kph.uni-mainz.de}}

\abstract{A systematic study of the $\eta$-$\eta'$ mixing in Large-$N_c$ chiral perturbation theory is presented~\cite{BMS}
with special emphasis on the role of the next-to-next-to-leading-order contributions in the combined
momentum, quark-mass, and $1/N_c$ expansions.
   At this order, loop corrections as well as OZI-rule-violating pieces need to be included.
   Mixing angles as well as pseudoscalar decay constants are discussed within this framework.
   The results are compared with recent phenomenological approaches.}
\FullConference{The 8th  International Workshop on Chiral Dynamics \\
         29 June 2015 - 03 July 2015\\
         Pisa, Italy}

\begin{document}

\section{Introduction}

   The pseudoscalar mesons $\eta$ and $\eta'$ represent an ideal laboratory for testing both (global) symmetries and
symmetry-breaking mechanisms in QCD at low energies.
   On the one hand, hadronic decays such as $\eta^{(\prime)}\to\pi\pi\pi$ and $\eta'\to\eta\pi\pi$ test our knowledge of low-energy effective field
theories (EFTs) and provide information on the light-quark masses.\footnote{For an overview of the main topics in
$\eta$ and $\eta'$ physics from both theoretical and experimental sides, see Refs.~\cite{Adlarson:2012bi} and references
therein.}
   On the other hand, electromagnetic decays such as $\eta^{(\prime)}\to\gamma^{(\ast)}\gamma^{(\ast)}$
proceed through the Adler-Bell-Jackiw
anomaly~\cite{Adler:1969gk}.
   In the case of virtual photons, the corresponding amplitudes reveal the electromagnetic structure in terms
of the transition form factors.

   Both $\eta$ and $\eta'$ mesons belong to the set of the lightest pseudoscalars of QCD, where the octet
$(\pi,K,\eta_8)$ comprises the (almost) Goldstone bosons resulting from a spontaneous chiral symmetry breaking
in the ground state of QCD from $\text{SU(3)}_L\times\text{SU(3)}_R$ to $\mbox{SU(3)}_V$.
   Chiral symmetry is explicitly broken by the quark masses, and SU(3) flavor symmetry is broken by the
fact that the strange quark is substantially heavier than the up and down quarks \cite{Leutwyler:2013wna}.
   By means of an orthogonal transformation with mixing angle $\theta$, the physical $\eta$ and $\eta'$ states, i.e., the mass eigenstates,
are usually expressed as linear combinations of the octet and singlet states $\eta_8$ and $\eta_1$ of the flavor symmetry \cite{Agashe:2014kda}.
   This can be easily represented in terms of a quadratic mass matrix, where the diagonal entries are given by
the squares of the octet and the singlet masses~\cite{Isgur:1976qg,Fritzsch:1976qc},
while the off-diagonal terms account for the SU(3)-symmetry-breaking
effects~\cite{Gasser:1984gg,Donoghue:1986wv,Gilman:1987ax,Schechter:1992iz,Bramon:1997va}.

   Because the flavor symmetry is broken, the $\eta_8$ and $\eta_1$ states are not degenerate in mass.
   Moreover, the $\mbox{U}(1)_A$ anomaly~\cite{'tHooft:1973jz} contributes only to the singlet mass.
   As a result of the mixing, the anomaly contribution is transferred to the $\eta$ and $\eta'$ states.
   A discussion of the $\eta$-$\eta'$ mixing in the framework of EFT
should consider both states as degrees of freedom and, for a perturbative treatment, the respective masses
should be small in comparison with a typical hadronic energy scale.
   Now, in the chiral limit, the $\eta'$ still remains massive.
   However, invoking the large-number-of-colors (L$N_c$) limit of QCD~\cite{'tHooft:1973jz},
the $\mbox{U}(1)_A$ anomaly disappears, and the assumption of an $\mbox{SU(3)}_V\times\mbox{U}(1)_V$
symmetry of the ground state implies that the singlet state is also massless.
   In other words, in the combined chiral and L$N_c$ limits,
QCD at low energies is expected to generate $(\pi,K,\eta,\eta')$ as the Goldstone bosons.
    This scenario is the starting point for Large-$N_c$ chiral perturbation theory (L$N_c$ChPT) as the EFT of QCD at low energies
including the singlet
field~\cite{Moussallam:1994xp,HerreraSiklody:1996pm,Leutwyler:1997yr,Kaiser:1998ds,HerreraSiklody:1998cr,Kaiser:2000gs,Borasoy:2004ua,Guo:2015xva},
which we will also refer to as U(3) effective theory.

   In the framework of L$N_c$ChPT, we perform a simultaneous expansion of (renormalized) Feynman diagrams
in terms of momenta $p$, quark masses $m$, and $1/N_c$.\footnote{It is understood that dimensionful variables need to be small in comparison
with an energy scale.}
   The three expansion variables are counted as small quantities, scaling as~\cite{HerreraSiklody:1996pm,Kaiser:1998ds,Kaiser:2000gs}
\begin{equation}\label{powerexp}
p=\mathcal{O}(\sqrt{\delta}),\ \ \ m=\mathcal{O}(\delta),\ \ \ 1/N_c=\mathcal{O}(\delta).
\end{equation}
Within the $\delta$ counting, we can establish a set of power-counting rules collected in Table~\ref{powercounting}.

\begin{table}[htdp]
\begin{center}
\begin{tabular}{|c|c|}
\hline
 & Order\\
 \hline
 Decay constant $F$ &  ${\cal O}(1/\sqrt{\delta})$\\
 Flavor trace & ${\cal O}(\delta)$\\
 $k$-meson vertex from ${\cal L}^{(i)}$ & ${\cal O}(\delta^{i+(k-2)/2})$\\
 Goldstone-boson propagator &  ${\cal O}(1/\delta)$\\
 Goldstone-boson loop & $\sim \frac{M^2}{F^2}\sim{\cal O}(\delta^2)$\\
\hline
\end{tabular}
\end{center}
\caption{Power counting of L$N_c$ChPT in terms of the parameter $\delta$.}
\label{powercounting}
\end{table}%

   According to Table~\ref{powercounting}, in this scheme a loop increases the order by $\delta^2$.
   Thus, any calculation in this framework at the loop-level needs then to be performed at next-to-next-to-leading order (NNLO).
   This order would demand the knowledge of the low-energy constants (LECs) of the order $p^4$ and of those of ${\cal O}(p^6)$ which are leading in $1/N_c$.
   For SU(3), the LECs at ${\cal O}(p^4)$ are well known, and information on some of the ${\cal O}(p^6)$ LECs is also rather well known \cite{Bijnens:2014lea}.
   With a suitable matching, one can translate the SU(3) values into the corresponding ones within the U(3) effective theory.
   While at ${\cal O}(p^4)$ the matching between SU(3) and U(3) can be appropriately performed, at ${\cal O}(p^6)$ the matching relations
are $1/N_c$ suppressed and for that reason neglected (see Ref.~\cite{BMS} for details).

   This is, however, not yet the full story.
   Due to the inclusion of the $1/N_c$ expansion, terms violating the Okubo-Zweig-Iizuka (OZI) rule appear perturbatively in our calculations.
   They will be accompanied by a set of LECs which are rather poorly known at order $\delta$ and basically unknown at higher orders.
   This poses a challenge for any prediction within this theory, and information from other sources, e.g.,
from a matching to a physical observable or a lattice simulation, will be required.
   At ${\cal O}(\delta^2)$, this can still be done and will allow us to study the $\eta$-$\eta'$ mixing systematically.

   Actually, one of the main features of this theory is the simultaneous treatment of the loop effects and of the OZI-rule-violating parameters,
with the advantage of comparing their relative strengths order by order.
   In section 3, we will find that OZI-rule-violating parameters can be more important than expected and cannot be naively neglected.

   The present work is part of a more exhaustive analysis of low-energy $\eta'$ dynamics within chiral effective field theory.
   Large-$N_c$ ChPT allows for a systematic study of $VP\gamma$ transitions
(where $P$ stands for $\eta$ and $\eta'$ and $V$ for vector mesons) as well as the meson-baryon interactions.
   When those studies are to be performed at the one-loop level, consistency would demand treating the $\eta$-$\eta'$ mixing also at
the one-loop level.

   In the present work we compute the $\eta$-$\eta'$ mixing within L$N_c$ChPT and discuss the pseudoscalar decay constants within the same framework.

\section{The effective Lagrangian and the $\eta$-$\eta'$ mixing}

    Applying the power counting of Eq.~(\ref{powerexp}) to the construction of the most general Lagrangian
including the $\eta_1$, the effective Lagrangian takes the form~\cite{Kaiser:2000gs}
\begin{equation}
\mathcal{L}_{\textrm{eff}}=\mathcal{L}^{(0)}+\mathcal{L}^{(1)}+\mathcal{L}^{(2)}+\ldots,
\end{equation}
   where the superscripts $(i)$ denote the order in $\delta$.
   The leading-order Lagrangian is given by \cite{Kaiser:2000gs}
\begin{equation}
\mathcal{L}^{(0)}=\frac{F^2}{4}\langle D_\mu U(D^\mu
U)^\dagger\rangle+\frac{F^2}{4}\langle\chi U^\dagger+U\chi^\dagger\rangle-\frac{1}{2}\tau(\psi+\theta_{QCD})^2,
\end{equation}
\noindent
where $\chi=2BM$, $M=\textrm{diag}(\hat{m},\hat{m},m_s)$, $\hat{m}=m_u=m_d$, and $D_\mu U=\partial_\mu U-ir_\mu U+iU l_\mu$ with $r_\mu=v_\mu+a_\mu$ and
$l_\mu=v_\mu-a_\mu$ (see Ref.~\cite{Scherer:2012zzd} for an introduction to ChPT).
   For simplicity, we discard the external fields $s$ and $p$, except for the quark-mass term.
   This Lagrangian contains 3 LECs, namely, $F$, $B$, and $\tau$.
   The singlet field $\eta_1$ is described by the dimensionless field $\psi$ in terms of $\mbox{det}\,U(x)=e^{i \psi(x)}$,
where $U$ is a unitary $3\times 3$ matrix.
   The pseudoscalar fields are encoded in
\begin{equation}
\label{states}
\phi(x)= \sum_{a=0}^8 \lambda_a \phi_a(x)=
\begin{pmatrix}
\pi^0+\frac{1}{ \sqrt{3} } \eta_8 + \frac{F}{3}\psi &  \sqrt{2}\pi^+ & \sqrt{2}K^+\\
\sqrt{2}\pi^- &  -\pi^0+\frac{1}{\sqrt{3}}\eta_8+\frac{F}{3}\psi & \sqrt{2}K^0\\
\sqrt{2}K^-  & \sqrt{2}\bar{K}^0  & -\frac{2}{\sqrt{3}}\eta_8+\frac{F}{3}\psi
\end{pmatrix}.
\end{equation}
   At NLO, the parts of $\mathcal{L}^{(1)}$ relevant for the $\eta$-$\eta'$ mixing read~\cite{HerreraSiklody:1996pm,Kaiser:2000gs}
\begin{align}
\label{L1}
\mathcal{L}^{(1)} &= L_5 \langle D_\mu U ^\dagger D^\mu U (\chi^\dagger U + U^\dagger \chi) \rangle
+ L_8 \langle \chi^\dagger U \chi^\dagger U  + U^\dagger \chi U^\dagger \chi \rangle\nonumber   \\
&\quad + \frac{F^2}{12} \Lambda_1 D_\mu \psi D^\mu \psi + i \frac{F^2}{12} \Lambda_2 \bar{\psi}
\langle \chi^\dagger U - U^\dagger \chi \rangle+\ldots,
\end{align}
\noindent
with $\bar{\psi}=\psi+\theta_{QCD}$
and $D_\mu\psi=\partial_\mu\psi$.\footnote{In the following, we set $\theta_{QCD}=0$.}
   The LECs $L_5$ and $L_8$ appear in the SU(3) sector as well and are rather well known \cite{Bijnens:2014lea},
while $\Lambda_1$ and $\Lambda_2$ belong to the singlet field and are poorly known.
   The ellipsis represents other terms which play no role in this calculation.

   Finally, at NNLO, the relevant pieces of $\mathcal{L}^{(2)}$ can be split into three different contributions
of the orders $N_c^{-1}p^2$, ${\cal O}(p^4)$, and ${\cal O}(N_c p^6)$, respectively~\cite{Fearing:1994ga,Bijnens:1999sh,Jiang:2014via}:
\begin{eqnarray}
\label{L2}
{\cal L}^{(2, N_c^{-1}p^2)} &=& -\frac{F^2}{4}v^{(2)}_2\bar{\psi}^2\langle \chi U^\dagger+U\chi^\dagger \rangle, \nonumber\\
{\cal L}^{(2, p^4)} &=& L_4 \langle D_\mu U ^\dagger D^\mu U \rangle\langle \chi^\dagger U + U^\dagger \chi  \rangle
+ L_6 \langle \chi^\dagger U + U^\dagger \chi  \rangle^2 + L_7 \langle \chi^\dagger U - U^\dagger \chi  \rangle^2 \nonumber\\
&& + i L_{18} D_\mu \psi \langle D^\mu U^\dagger \chi - D^\mu U \chi^\dagger \rangle
+ i L_{25} \bar{\psi} \langle U^\dagger \chi U^\dagger \chi - \chi^\dagger U \chi^\dagger U   \rangle+\ldots,\\ \nonumber
{\cal L}^{(2,N_cp^6)} &= &C_{12} \langle \chi_{+}h_{\mu\nu}h^{\mu\nu} \rangle
+ C_{14} \langle u_{\mu}u^{\mu}\chi^2_{+}\rangle+C_{17} \langle \chi_{+}u_{\mu}\chi_{+}u^{\mu}\rangle  + C_{19} \langle\chi^3_{+}\rangle+C_{31} \langle \chi^2_{-}\chi_{+}\rangle
+\ldots,
\end{eqnarray}
where $h_{\mu\nu}=\nabla_\mu u_\nu+\nabla_\nu u_\mu$, $u_{\mu}=i\left\{u^\dagger(\partial_\mu-ir_\mu)u-u(\partial_\mu-il_\mu)u^\dagger\right\}$,
and $\chi_\pm=u^\dagger\chi u^\dagger\pm u\chi^\dagger u$.
   The LECs $L_4$, $L_6$, and $L_7$ originate from the SU(3) sector and are known \cite{Bijnens:2014lea}.
   Because of the two flavor traces, they are suppressed by $1/N_c$ with respect to $L_5$ and $L_8$ of Eq.~(\ref{L1}).
   Similarly, $L_{18}$ and $L_{25}$ are suppressed by $1/N_c$ with respect to $\frac{F^2}{12} \Lambda_1$ and $\frac{F^2}{12} \Lambda_2$,
and at present they are unknown.
   The naive counting would suggest them to be of the order of their homologous LECs $L_4$, $L_6$, and $L_7$.
   Moreover, $C_{12}$, $C_{14}$, $C_{17}$, $C_{19}$, and $C_{31}$ are the leading coefficients in the $1/N_c$ expansion of the ${\cal O}(p^6)$ Lagrangian in SU(3)
and are poorly known.
   Finally, in the current discussion $v^{(2)}_2$ is set to zero.

   The above Lagrangians describe the dynamics in terms of the fields collected in Eq.~(\ref{states}), in particular, the $\eta_8$ and $\eta_1$ fields.
   The physical $\eta$ and $\eta'$ states observed experimentally are, however, different.
   To relate the mathematical states with their physical counterparts, we define a mixing pattern among the
corresponding fields.
   For that purpose, we introduce an effective Lagrangian responsible for the mixing in terms of $\eta_{BC}^T \equiv (\eta_8,\eta_1)$,
\begin{equation}
\label{Lmixing1}
\mathcal{L}=\partial_{\mu}\partial_{\nu}\eta^T_{BC}\mathcal{C}\partial^{\mu}\partial^{\nu}\eta_{BC}+
\frac{1}{2}\partial_{\mu}\eta^T_{BC}\mathcal{K}_0\partial^{\mu}\eta_{BC}-\frac{1}{2}\eta^T_{BC}\mathcal{M}^2\eta_{BC},
\end{equation}
with
\begin{equation}
\mathcal{C}=
\begin{pmatrix} c_8&c_{81} \\
                                 c_{81} &  c_1\\
 \end{pmatrix}
\textrm{,} \quad
\mathcal{K}_0=
\begin{pmatrix}
1+\delta^{(1)}_8+\tilde{\delta}^{(2)}_8 &\delta^{(1)}_{81}+\tilde{\delta}^{(2)}_{81} \\
\delta^{(1)}_{81}+\tilde{\delta}^{(2)}_{81} &  1+\delta^{(1)}_1+\tilde{\delta}^{(2)}_1\\
\end{pmatrix}
\textrm{,}\quad \textrm{and} \quad
\mathcal{M}^2= \begin{pmatrix}
M^2_8 & M^2_{81} \\
M^2_{81}  &  M^2_{1} \\
\end{pmatrix}.
\end{equation}
   To get the ''physical'' fields $\eta_P\equiv (\eta, \eta')^T$, Eq.~(\ref{Lmixing1}) needs to be diagonalized
(a detailed discussion of this procedure can be found in Ref.~\cite{BMS}).
   After the field redefinition $\eta_{BC}= Z_C\cdot\eta_{B}$ to remove the higher-derivative terms, the Lagrangian takes the form
\begin{align}
\mathcal{L}=\frac{1}{2}\partial_{\mu}\eta^T_B\mathcal{K}\partial^{\mu}\eta_B-\frac{1}{2}\eta^T_B\mathcal{M}^2\eta_B,
\label{Lmixing}
\end{align}
where the expressions for the new kinetic matrix $\mathcal{K}$ and for $Z_C$ can be found in Ref.~\cite{BMS}.
   We then perform two additional transformations, $Z^{1/2}$ and $R$, to diagonalize the Lagrangian in Eq.~(\ref{Lmixing}),
and the full transformation for the states reads
\begin{equation}
\eta_{BC}= Z_C\cdot {Z^{1/2}}^T\cdot R^T \cdot \eta_P,
\end{equation}
where the matrices $Z_C$, $Z^{1/2}$, and $R$ are given in Ref.~\cite{BMS}. The mass matrix $\mathcal{M}^2$ in Eq.~(\ref{Lmixing}) is related to the diagonal mass matrix
$\mathcal{M_D}^2$ of the physical $\eta$ and $\eta'$ fields via
\begin{equation}\label{mdiag}
\mathcal{M_D}^2 = R\cdot\hat{\mathcal{M}}^2\cdot R^T,\ \ \ \text{with}\ \ \hat{\mathcal{M}}^2=Z^{1/2}\cdot \mathcal{M}^2 \cdot{Z^{1/2}}^T.
\end{equation}
\noindent
   Equation~(\ref{mdiag}) yields the desired mixing angle $\theta^{(2)}$ in terms of the calculated matrix
elements $\hat{M}_1^2$, $\hat{M}_8^2$, and $\hat{M}_{81}^2$ and the physical masses $M_\eta^2$ and $M_{\eta'}^2$:
\begin{align}
\label{eq:masses1}
\hat{M}^2_8&= M_{\eta}^2 \cos^2 \theta^{(2)} + M_{\eta'}^2 \sin^2 \theta^{(2)}, \\
\label{eq:masses2}
\hat{M}^2_1&= M_{\eta}^2 \sin^2 \theta^{(2)}+ M_{\eta'}^2 \cos^2 \theta^{(2)},  \\
\label{eq:masses3}
\hat{M}^2_{81}&=(M_{\eta}^2 - M_{\eta'}^2) \sin \theta^{(2)} \cos \theta^{(2)},
\end{align}
where the superscript (2) refers to the fact that the mixing angle is calculated up to
and including order $\delta^2$.
   As an example, the third equation can be solved to yield
\begin{equation}
\textrm{sin}2\theta^{(2)}=\frac{2\hat{M}^2_{81}}{M^2_{\eta'}-M^2_{\eta}}\, .
\end{equation}
   Moreover, we obtain
\begin{equation}
\label{eq:mass_relations}
M^2_{\eta'}+M^2_{\eta}=\hat{M}^2_8+\hat{M}^2_1,\ \ \ M^2_{\eta'}-M^2_{\eta}=\sqrt{(\hat{M}^2_8-\hat{M}^2_1)^2+4 \hat{M}^4_{81}}.
\end{equation}
    The quantities $\hat{M}_1^2$, $\hat{M}_8^2$, and $\hat{M}_{81}^2$ are complicated expressions that can, in principle, be perturbatively
evaluated at any given order:
\begin{align*}
\hat{M}^2_8&=\stackrel{\circ}{M^2_8}+\Delta {M^2_8}^{(1)}+\Delta {M^2_8}^{(2)},\\ \nonumber
\hat{M}^2_1&=M^2_0+\stackrel{\circ}{M^2_1}+\Delta {M^2_1}^{(1)}+\Delta {M^2_1}^{(2)},\\ \nonumber
\hat{M}^2_{81}&=\stackrel{\circ}{M^2_{81}}+\Delta {M^2_{81}}^{(1)}+\Delta {M^2_{81}}^{(2)}. \nonumber
\end{align*}
   Here, the $\stackrel{\circ}{M^2_{i}}$ refer to the leading-order masses.
   The full expressions for $\Delta {M^2_{i}}^{(1,2)}$ are provided in Ref.~\cite{BMS}.

\section{Preliminary results}
\subsection{Remarks on convergence}

   Using L$N_c$ChPT with the LECs of tables 3 and 4 of Ref.~\cite{Bijnens:2014lea} in combination with the SU(3)--U(3) matching,
we obtain for the ratio $F_K/F_\pi$:
\begin{equation}
\label{FkFpi}
F_K/F_\pi \simeq \underbrace{1}_{\textrm{LO}}+\underbrace{0.15}_{\textrm{NLO}}+\underbrace{0.03}_{\textrm{NNLO}}\, ,
\quad \, \textrm{ where }\quad  0.03 \simeq \underbrace{0.05}_{\textrm{loop}}- \underbrace{0.01}_{C_i} -\underbrace{0.02}_{L_i L_j}.
\end{equation}
   The result displays a nice convergence pattern both with respect to the different orders (cf.~Refs.~\cite{Ecker:2010nc,Ecker:2013pba})
and also within the NNLO, namely, the different contributions show a hierarchy loop $> C_i \sim L_iL_j$.
   A similar pattern is also observed within SU(3) ChPT~\cite{Bijnens:2014lea},
\begin{equation}
F_K/F_\pi \simeq 1+0.18+0.02,
\end{equation}
where the three terms represent LO, NLO, and NNLO, respectively.

   On the other hand, the kaon mass represents a somewhat more involved case.
   Using the same prescription as in Eq.~(\ref{FkFpi}), we find within L$N_c$ChPT the following perturbative pattern
at $\mu=0.77$~GeV:
\begin{equation}
M_K^2/M_{K,\textrm{phys}}^2 \simeq 1.26 + 0.01 - 0.27, \quad \textrm{where}\quad
-0.27 \simeq -\underbrace{0.15}_{\textrm{loop}}-\underbrace{0.14}_{C_i}+\underbrace{0.03}_{L_4,L_6}-\underbrace{0.01}_{L_i L_j}\, .
\end{equation}
   Loops are now comparable with the $C_i$, although in SU(3) they belong to different orders.
   The NNLO pieces read at $\mu=0.77$~GeV:
\begin{eqnarray}
-\underbrace{0.15}_{\textrm{loop}} = -\underbrace{0.10}_{\eta'}-\underbrace{0.05}_{\pi,K,\eta}\,
\quad \textrm{and} \quad -\underbrace{0.14}_{C_i} = -\underbrace{0.29}_{C_{19}}+\underbrace{0.15}_{C_{12,14,17,31}}.
\end{eqnarray}
   The $\eta'$-loop contribution is large in comparison with the other loops, which is an interesting feature for the kaon mass.
   At NNLO the $C_{19}$  contribution dominates. For that, we did not find a simple explanation beyond the fact that already for SU(3)
the numerical value for that LEC is quite large~\cite{Bijnens:2014lea}.
   Errors for the $C_i$ are not provided, so it could still be that the large impact we observe here
is made smaller with an eventual treatment of the errors of the LECs.
   In SU(3), the expansion of the same observable reads~\cite{Bijnens:2014lea}
\begin{equation}
M_K^2/M_{K,\textrm{phys}}^2 \simeq 1.11 -0.07 -0.04\, .
\end{equation}

   We conclude from the above discussion that even though some of the SU(3) LECs are known up to and including ${\cal O}(p^6)$,
the lack of errors on the one hand and neglecting the matching between SU(3) and U(3) for the $C_i$
may lead to unexpected results.
   Numerical analyses are, then, difficult to perform, and we need a strategy to fill in all the required input
parameters, and to discuss their respective relevance.

\subsection{Numerical results for the $\eta$-$\eta^\prime$ mixing angle $\theta$}

\begin{figure*}[htb]
\begin{center}
    \centering
    \includegraphics[width=0.5\textwidth]{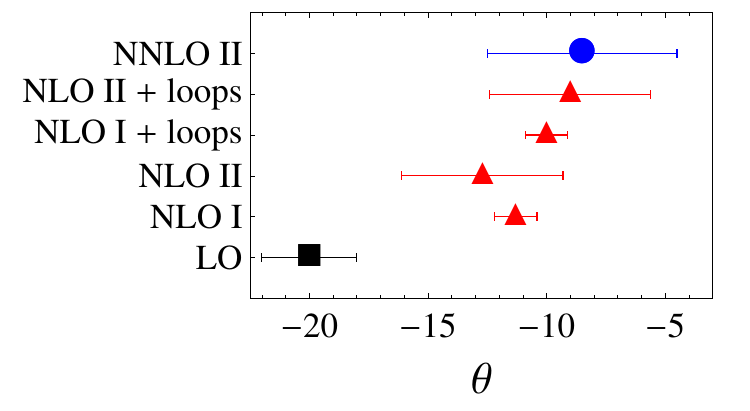}
\caption{Mixing angle $\theta$ of the $\eta$-$\eta^\prime$ system in L$N_c$ChPT at different orders in $\delta$.}
\label{fig:mixingangle}
\end{center}
\end{figure*}

   At leading order, the mixing angle is given by
\begin{equation}
\label{eq:sin2theta0}
\textrm{sin}\:2\theta^{(0)}=
\frac{-4\sqrt{2}(\stackrel{\circ}{M^2_{K}} - \stackrel{\circ}{M^2_{\pi}})}{\sqrt{(2\stackrel{\circ}{M^2_K}-2\stackrel{\circ}{M^2_{\pi}} - M_0^2)^2
+32 (\stackrel{\circ}{M^2_{K}} - \stackrel{\circ}{M^2_{\pi}})^2 }}\,.
\end{equation}
   Using Eq.~(\ref{eq:mass_relations}) for $M_0^2$ at leading order and replacing the
leading-order masses by the physical masses,
\begin{equation}
M_0^2 =\frac{(M_{\eta'}^2-M_{\pi}^2)(M_{\eta'}^2- 2 M_K^2 + M_{\pi}^2)}{M_{\eta'}^2 -\frac43 M_K^2 + \frac{1}{3}M_{\pi}^2}=(0.820 \textrm{ GeV})^2\, ,
\end{equation}
one obtains from Eq.~(\ref{eq:sin2theta0}) the value $\theta^{(0)}=-19.6^\circ$ (black square in Fig.~\ref{fig:mixingangle}).

   At NLO, things are more involved.
   With the caveats presented above in mind, we make use of two different strategies to perform our numerical studies of the mixing angle.
   Strategy NLO~I consists of determining the unknown LECs by calculating a set of suitable observables to match our expressions with the physical
counterparts.
   Strategy NLO~II employs the LECs obtained in Ref.~\cite{Bijnens:2014lea} after the matching to U(3) and uses extra physical information
to obtain values for the OZI-rule-violating parameters.
   The results are shown as red triangles in Fig.~\ref{fig:mixingangle}.
   In the next step, we include the loop contributions within the two scenarios.
   The new results are shown in Fig.~\ref{fig:mixingangle} under the labels NLO I + loop and NLO II + loop, respectively.

   At NNLO, our expressions depend on too many unknown LECs.
   However, they appear in particular combinations, and the power counting of our theory may provide a guidance towards their determination.
   We have investigated the dependence of various observables on unknown LECs.
   For example, setting $\Lambda_2=0$, $M_\eta^2$ can be written as $M_\eta^2(\Lambda_1,\tilde{L})$,
where $\tilde{L}=L_{18}+2L_{25}$.
   One finds that $M_\eta^2(\Lambda_1, \tilde{L})$ has a quadratic dependence on $\Lambda_1$, while it depends only linearly on $\tilde{L}$.
   Both $\Lambda_1$ and $\tilde{L}$ are unknown but we can impose that $M_\eta^2(\Lambda_1, \tilde{L})=M_{\eta,\textrm{phys}}^2$.
   Solving the quadratic equation for $\Lambda_1$ and demanding that all the parameters should be real, we find a minimal bound for $\tilde{L}$
which cannot be surpassed for $\text{Im}(\Lambda_1)=0$.
   At the same time, taking this minimal value for $\tilde{L}$ and solving the equation for $M_\eta^2$, we can find a value for $\Lambda_1$.
   For values of $\tilde{L}$ larger than the minimal bound, we obtain two solutions for
$\Lambda_1$ that range basically from $-1$ to $+1$.

   On the other hand, $\tilde{L}$ cannot be arbitrarily large.
   In order to keep $M_\eta^2$ at its physical value, $\Lambda_1$ would have to increase to compensate for $\tilde{L}$,
and the perturbative expansion in terms of $1/N_c$ would be spoiled.
   Varying both $\Lambda_1$ and $\tilde{L}$ within the ranges just defined will provide us with an estimate of our input errors.
   With this range of parameters we can evaluate the mixing angle $\theta$ at NNLO.
   The result is shown in Fig.~\ref{fig:mixingangle} as a blue circle.
   It should be pointed out that other LECs are set to zero, and no error from that consideration
is taken into account in this preliminary investigation.

\subsection{Decay constants of the $\eta$-$\eta'$ system in the octet-singlet basis}

\begin{figure*}[htb]
\begin{center}
    \centering
    \includegraphics[width=0.4\textwidth]{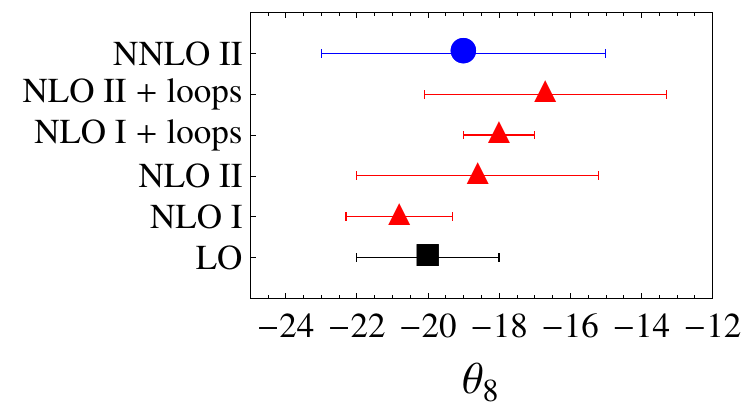}
    \hspace{0.0cm}\includegraphics[trim=2.5cm 0cm 0cm 0cm, clip=true,width=0.267\textwidth]{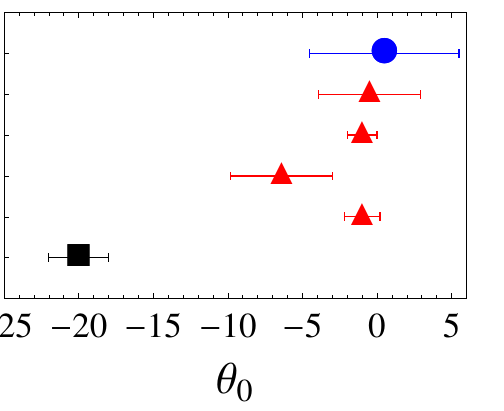}\\
    \includegraphics[width=0.4\textwidth]{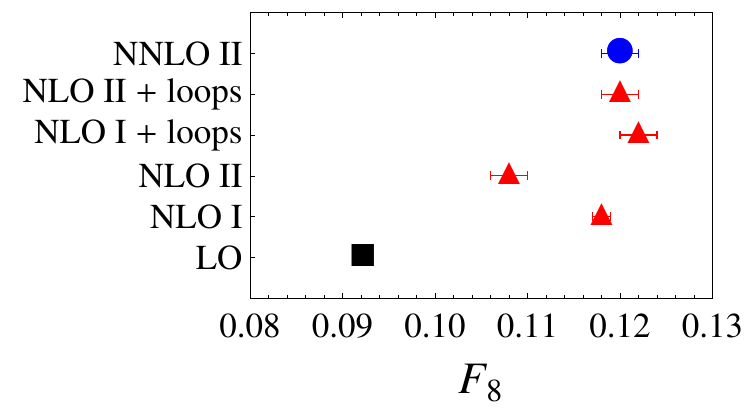}
    \hspace{0.0cm}\includegraphics[trim=2.5cm 0cm 0cm 0cm, clip=true,width=0.267\textwidth]{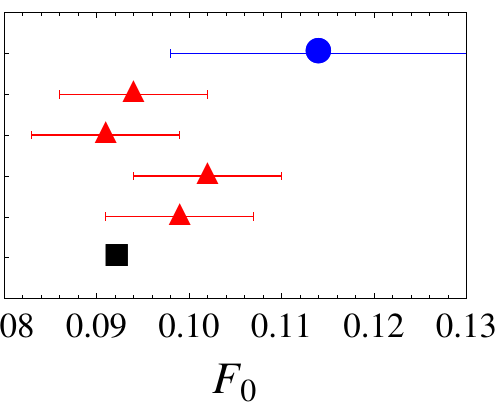}\\
\caption{Mixing parameters of the $\eta$-$\eta^\prime$ system in the octet-singlet basis at different orders.}
\label{fig:mixingparam}
\end{center}
\end{figure*}

\begin{figure*}[htb]
\begin{center}
    \centering
    \includegraphics[width=0.4\textwidth]{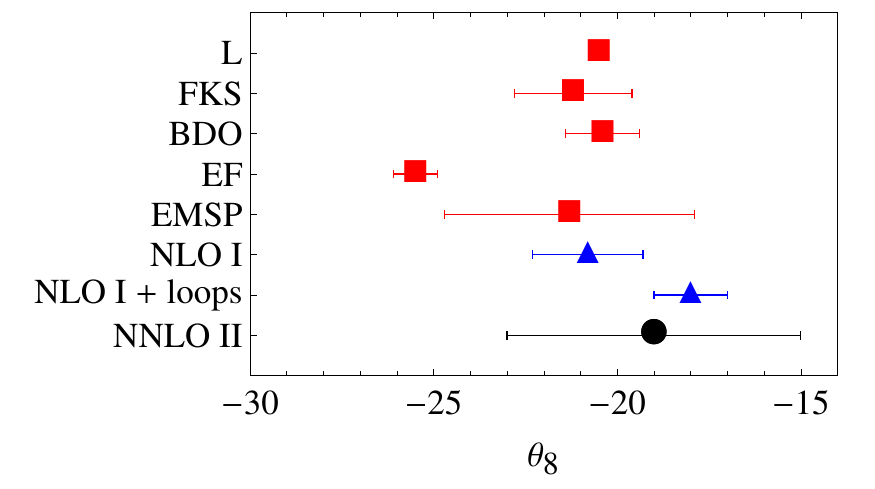}
    \hspace{0.0cm}\includegraphics[trim=2.5cm 0cm 0cm 0cm, clip=true,width=0.286\textwidth]{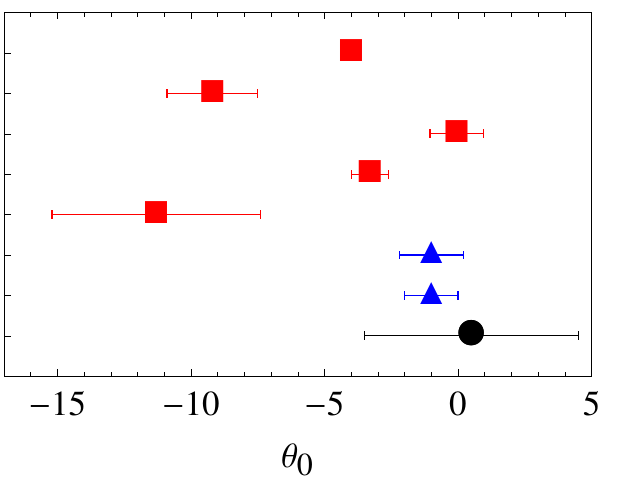}\\
\caption{Mixing angles of the $\eta$-$\eta^\prime$ system in the octet-singlet basis from different references (red squares L~\cite{Leutwyler:1997yr},
FKS~\cite{Feldmann:1998vh}, BDO~\cite{Benayoun:1999au}, EF~\cite{Escribano:2005qq}, EMSP~\cite{Escribano:2013kba}),
compared with our NLO (blue triangles) and NNLO (black circle) results. See text for details.}
\label{fig:mixing80}
\end{center}
\end{figure*}

   We also calculate the axial-vector decay constants of the $\eta$-$\eta'$ system at NNLO and determine the mixing parameters
$F_8$, $F_0$, $\theta_8$, and $\theta_0$ of the so-called
two-angles scheme~\cite{Leutwyler:1997yr,Feldmann:1998vh,Feldmann:1998sh,Benayoun:1999au,Escribano:2005qq,Escribano:2010wt,Escribano:2013kba}.
   The decay constants are defined via the matrix element of the axial-vector current operator,
\begin{equation}
 \bra{0}A^{a}_\mu(0)\ket{P(p)}=iF^{a}_P p_\mu,
\label{f80}
\end{equation}
where $a=8,0$ and $P=\eta,\eta'$.
   Since both mesons have octet and singlet components, Eq.~(\ref{f80}) defines four independent decay constants, $F^a_P$.
   We parameterize them according to the convention in Ref.~\cite{Leutwyler:1997yr},
\begin{eqnarray}
F^a_P=\begin{pmatrix}
 F^8_\eta & F^0_\eta\\
F^8_{\eta'} & F^0_{\eta'}
\end{pmatrix}
=\begin{pmatrix}
F_8\cos\theta_8 & -F_0\sin\theta_0 \\
F_8\sin\theta_8 & F_0\cos\theta_0
\end{pmatrix}.
\end{eqnarray}
   The angles $\theta_8$ and $\theta_0$ and the constants $F_8$ and $F_0$ are given by
\begin{align}\label{angles}
\tan\theta_8=\frac{F^8_{\eta'}}{F^8_{\eta}},\quad\quad
\tan\theta_0=-\frac{F^0_\eta}{F^0_{\eta'}},
\end{align}
\begin{align}\label{F80}
F_8=\sqrt{(F^8_\eta)^2+(F^8_{\eta'})^2},\quad\quad
F_0=\sqrt{(F^0_\eta)^2+(F^0_{\eta'})^2}.
\end{align}
Figure~\ref{fig:mixingparam} represents a collection of our results in the octet-singlet basis for $\theta_{8,0}$ and $F_{8,0}$.
Figure~\ref{fig:mixing80} shows a comparison between our preliminary results and several representative numbers found in the literature.

\section{Conclusions and Outlook}
   In this work we have presented our preliminary results of the mixing parameters of the $\eta$-$\eta'$ system.
   Our calculation has been performed within L$N_c$ChPT, an effective field theory at low energies based on a simultaneous expansion in terms of
momenta, quark masses, and $1/N_c$.
   Including loop effects in this framework requires calculations at NNLO.
   Due to the $1/N_c$ expansion, the singlet field $\eta_1$ is included systematically in our calculations which, in turn, allow for the
study of the OZI-rule-violating terms.
   The main results of this preliminary work are collected in Figs.~\ref{fig:mixingangle}, \ref{fig:mixingparam}, and \ref{fig:mixing80}.
   Figure \ref{fig:mixing80} provides a comparison of the two mixing angles $\theta_8$ and $\theta_0$ with phenomenological determinations
found in the literature.
   The numerical analysis has to deal with a proliferation of unknown LECs.
   We have developed two different strategies to address this problem.
   Other numerical scenarios with a detailed discussion of loop effects and different determinations of LECs can be
found in Ref.~\cite{BMS}.

P.~M.~wants to thank O.~Cata, G.~Ecker, R.~Escribano, S.~Gonz\'alez-Sol\'is, M.~Kolesar, B.~Moussallam, and P.~Sanchez-Puertas
for useful discussions during the conference on the topic of this talk.
   Additionaly he wants to thank L.~Marcucci and M.~Viviani and the rest of the organizers of the 8th International Workshop on Chiral Dynamics for encouragement and support.
This work was supported by the Deutsche Forschungsgemeinschaft through the Collaborative Research Center ``The Low-Energy Frontier of the Standard Model'' (SFB 1044).

\end{document}